\documentclass[aps,prl,print,tightenlines,twocolumn,floatfix,amsmath,amssymb,superscriptaddress,altaffillsymbol,nofootinbib]{revtex4-2}

\usepackage{graphicx}
\usepackage{bm}
\usepackage{hyperref}
\usepackage{ulem}
\usepackage[utf8]{inputenx}
\usepackage[OT1]{fontenc}
\usepackage[final]{microtype}
\usepackage[x11names]{xcolor}
\usepackage{epstopdf}
\usepackage{dcolumn}
\usepackage[nameinlink,capitalise]{cleveref}

\usepackage{svg}
\svgsetup{inkscapelatex=false}
\DeclareMathAlphabet{\mathsfit}{T1}{\sfdefault}{\mddefault}{\sldefault}
\SetMathAlphabet{\mathsfit}{bold}{T1}{\sfdefault}{\bfdefault}{\sldefault}
\providecommand\bnabla{\boldsymbol{\nabla}}
\providecommand\bcdot{\boldsymbol{\cdot}}
\DeclareMathAlphabet\mathsfbi            {OT1}{cmss}{m}{sl}
\IfFileExists{t1phv.fd}
{\DeclareTextFontCommand\textsfbi{\usefont{T1}{phv}{b}{it}}
	\DeclareMathAlphabet\mathsfbi            {T1}{phv}{b}{it}
}{}
\IfFileExists{ot1phv.fd}
{\DeclareTextFontCommand\textsfbi{\usefont{OT1}{phv}{b}{it}}
	\DeclareMathAlphabet\mathsfbi            {OT1}{phv}{b}{it}
}{}

\hypersetup{
    colorlinks=true,
    linkcolor=blue!85!black,
    citecolor=blue!85!black,
    urlcolor=blue!85!black,
    pdfauthor={V. Dzanic, S. P. Thampi, J. M. Yeomans},
    pdftitle={Bridging Elastic and Active Turbulence}
}

\newcommand{\ju}[1]{\textcolor{red}{#1}}

\begin{document}

% Old Title
% \title{Bridging Elastic and Active Turbulence}

% New Title
\title{Connecting the Hydrodynamics of Polymeric Fluids and Active Nematics}
%\title{Connecting the Hydrodynamics of Polymer\su{ic \rv{F}luid}s and \bluestrike{Deformable} Active Nematics}

\author{Vedad Dzanic}
\email{vedad.dzanic@physics.ox.ac.uk}
\affiliation{
School of Mechanical, Medical, and Process Engineering, \\
Queensland University of Technology, Brisbane, QLD 4001, Australia
}
\affiliation{
The Rudolf Peierls center for Theoretical Physics, Department of Physics, \\
University of Oxford, Parks Road, Oxford OX1 3PU, United Kingdom
}

\author{Sumesh P.~Thampi}
\email{sumesh@iitm.ac.in}
\affiliation{
Department of Chemical Engineering, \\
Indian Institute of Technology Madras, Chennai 600036, India
}

\author{Rahil N. Valani}
\email{rahil.valani@physics.ox.ac.uk}
\affiliation{
The Rudolf Peierls center for Theoretical Physics, Department of Physics, \\
University of Oxford, Parks Road, Oxford OX1 3PU, United Kingdom
}

\author{Julia M.~Yeomans}
\email{julia.yeomans@physics.ox.ac.uk}
\affiliation{
The Rudolf Peierls center for Theoretical Physics, Department of Physics, \\
University of Oxford, Parks Road, Oxford OX1 3PU, United Kingdom
}

\date{\today}

\begin{abstract}
We establish a correspondence between the continuum descriptions of polymer solutions and active nematics, two seemingly disparate but well-studied complex fluids. Through a mapping of the governing equations, we  demonstrate that polymeric fluids can be interpreted as deformable, contractile, high-temperature active nematics, with an effective activity controlled by polymer extension. Using numerical simulations under Kolmogorov flow, we construct a phase diagram connecting the nematic and polymer limits, highlighting how their one-and two-dimensional instabilities depend on deformability. This
encompasses both a flow-suppressed regime, in which active stresses oppose the imposed flow, and a transverse instability that generates secondary flows and leads towards elastic turbulence.  %\cyanstrike{At high activity, a balance between active and imposed Kolmogorov stresses suppresses the flow, yielding weak polymer stretching and orientational order.} 
The transverse instability is facilitated by  %\bluestrike{shown to require} 
spatial gradients in activity, which arise naturally in extensible polymers through variations in chain extension. 
%\bluestrike{, but require external imposition in inextensible active nematics}. 
We further show that the transition into the well-known traveling arrowhead structures in elastic turbulence is marked by the emergence of charge $\pm 1/2$ topological defects in the polymer director field, indicating emergent nematic ordering of stretched polymers.
Our results establish a unified framework connecting polymer and active matter hydrodynamics, with implications for low–Reynolds-number turbulence, active polymer systems, and the mechanics of deformable living matter.

\end{abstract}

\maketitle

\section{Introduction}\label{introduction}

Even in the absence of inertia, intricate 
flows can emerge in complex fluids through the dynamics of their microstructural constituents. One example occurs when long-chain, flexible polymers are added to a background fluid flow. The polymers are stretched and aligned by flow gradients, and their relaxation generates contractile elastic stress contributions that can drive the flow into a vortex lattice or into a chaotic regime, coined elastic turbulence \cite{Groisman_2001,Steinberg_REVIEW}. A second example of such inertialess
flows arises in dense active nematic systems, which comprise self-driven 
% \redstrike{, disc-like or rod-like} 
particles that continuously transform energy into motion. Examples include bacteria \cite{Bact_Turb,Dombrowski,Bact_turb_SciAdv}, microtubules driven by molecular motors \cite{Sanchez_Microtub,Guillamat2017}, and epithelial cells \cite{Soft_matt_Cell,PRL_Cell_Turb,Nature_Top_Def_Cell}. Because the particles are self-propelled, active nematics exert dipolar forces \cite{LSA_Ramaswamy}, and the resulting stresses drive hydrodynamic instabilities that generate complex flows, for example %\cyanstrike{lines} 
a lattice of velocity rolls or a chaotic state, 
% \redstrike{flows}
 commonly referred to as active turbulence \cite{ACTIVE_TURB_ANNUAL_REV,Thampi2014}. In this work, we establish a correspondence between the continuum descriptions of these two seemingly disparate complex fluids. 
%\ju{We identify dimensionless variables that relate the physics of active nematics, deformable active nematics and dilute polymer suspensions, and hence connect their hydrodynamics and instabilities, and identify an activity-driven jamming transition.}}

\begin{figure*}[t]
	\centering
    \includegraphics[width=\textwidth]{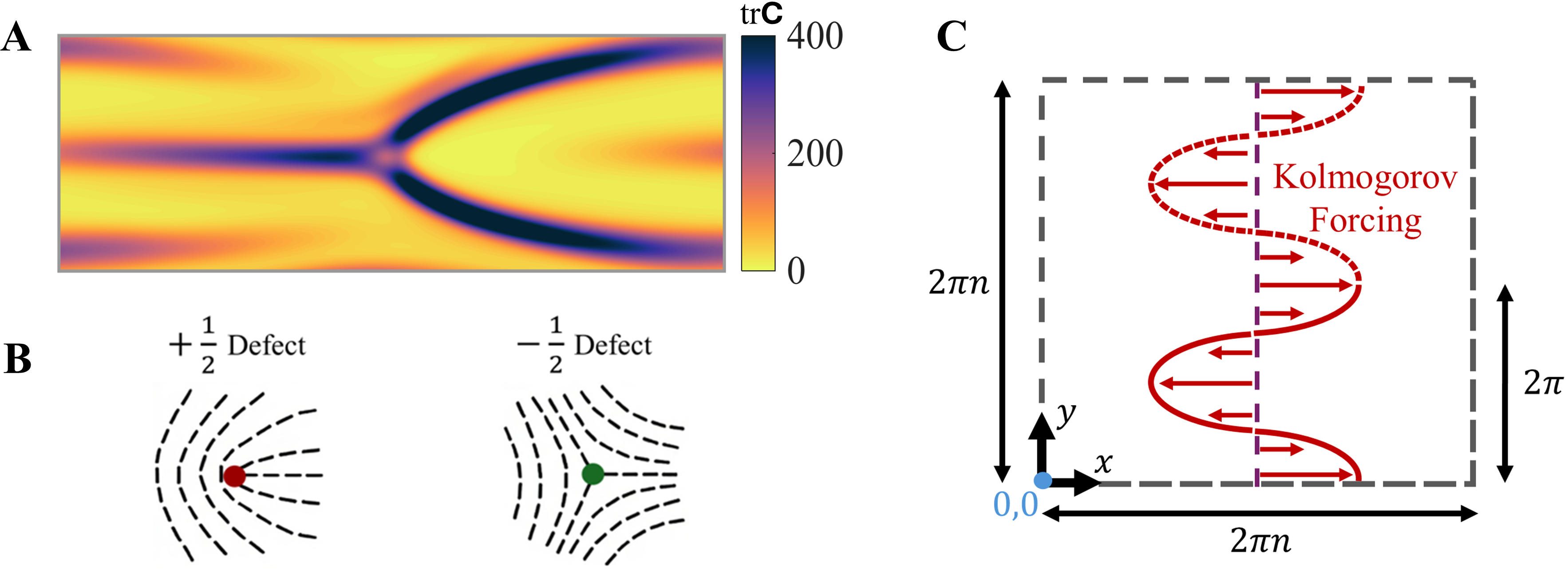}%width=0.95\textwidth
	\caption{
    (\textbf{A}) A coherent arrowhead structure in a polymeric fluid, visualized through the spatial distribution of \(\mathrm{tr}(\mathsfbi C)\), which characterizes local polymer extension. (\textbf{B}) Schematic illustrations of the director-field configurations around (+1/2) (left, red dot) and (-1/2) (right, green dot) topological defects, exhibiting characteristic comet-like and trefoil-like patterns, respectively.  (\textbf{C}) Schematic of the fully periodic domain used to impose  Kolmogorov flow  with sinusoidal forcing (shown in red) applied along the $x$-axis. The forcing wavelength is $2\pi$ and the number of wavelengths is controlled by the integer $n$. }
	\label{introfig}
\end{figure*}

%\bluestrike{The classical understanding of turbulence is rooted in the dominance of inertial effects, characterized by high Reynolds numbers $Re\gg1$, where strong, externally imposed forces give rise to chaotic flow dynamics marked by a cascade of energy from large to small scales. However, in recent decades, a new class of turbulence has emerged under inertialess conditions (i.e., $Re\ll1$), where chaotic dynamics are instead sustained through internal stress contributions arising from microstructural agents. One such example can occur if  long-chain, flexible polymers are added to a background fluid. The polymers are stretched and aligned by flow gradients, and their relaxation generates contractile elastic stress contributions that can transition the flow into a chaotic regime, coined \textit{elastic turbulence} \cite{Groisman_2001,Steinberg_REVIEW}. E}
The two systems have traditionally been studied from different perspectives. In polymeric fluids, elastic turbulence is distinguished by increased flow resistance \cite{Groisman_2004,Browne_ET}, enhanced mixing \cite{Groisman_Mix,Browne_PNAS} and random flow fluctuations characterized by a broadband spectrum of spatial and temporal frequencies \cite{Groisman_2004,Datta_Pers,Singh_ET}. Recent studies \cite{Page_PRL,Morozov_Coherent} have further revealed the presence of exact coherent structures in parallel shear flows. These take the form of traveling arrowhead-like patterns (Fig.~\ref{introfig}A), arising from a center-mode instability \cite{Lewy_Kerswell_2025}, highlighting organized features within otherwise chaotic velocity fields.

%\indent \bluestrike{Similar inertialess chaotic flows emerge in dense active nematic systems, which comprise self-driven, disc-like or rod-like particles that continuously transform energy into motion. Examples include bacteria \cite{Bact_Turb,Dombrowski,Bact_turb_SciAdv}, microtubules driven by molecular motors \cite{Sanchez_Microtub,Guillamat2017}, and epithelial cells \cite{Soft_matt_Cell,PRL_Cell_Turb,Nature_Top_Def_Cell}. Because the particles are self-propelled, active nematics exert dipolar forces \cite{LSA_Ramaswamy}, and the resulting stresses drive a hydrodynamic instability that leads to a state commonly referred to as \textit{active turbulence} \cite{ACTIVE_TURB_ANNUAL_REV,Thampi2014}.The active turbulent state is} 
In active nematics,  active flows are initiated by the instability of nematic order to  deformations in the nematic director field \cite{Ramaswamy_Splay,LSA_Ramaswamy}. 
%\redstrike{Spatial gradients in both the direction and magnitude of the nematic order parameter generate the forces that drive the flow.} 
A defining feature of 
%\redstrike{active nematic} 
the flows is the appearance of the distinctive director patterns shown in Fig.~\ref{introfig}B,  termed topological defects, and labeled by their topological charge $\pm 1/2$. At high activities, there is a chaotic state distinguished by high vorticity and the spontaneous creation, motion, and annihilation of topological defects \cite{Thampi2014,Doostmohammadi18,Giomi14b,ACTIVE_TURB_ANNUAL_REV}.

Despite the different properties emphasized in the literature, the two systems share several important similarities. Both exhibit
%\bluestrike{In addition to both exhibiting}
low-Reynolds-number 
% \bluestrike{chaos}
fluid flow instabilities and complex flow patterns \cite{Steinberg_REVIEW,ACTIVE_TURB_ANNUAL_REV}. In both systems, the microstructural constituents
% \bluestrike{, active nematics and extended polymers share several other properties: they} 
act as contractile hydrodynamic force dipoles and the the continuum equations of motion describing their microstructural dynamics are written in terms of rank-2 order tensors \cite{Alves_Rev,Beris1994}. Moreover, chaotic flows in the two systems can exhibit similar scaling properties \cite{Steinberg_Scaling,MOROZOV_PNAS,Alert2020}. %
%act as contractile hydrodynamic force dipoles \ju{and the} the continuum equations of motion describing their microstructural dynamics are written in terms of rank-2 order tensors \cite{Alves_Rev,Beris1994}. \vedcomment{Moreover, the two systems typically exhibit similar scaling properties \cite{Steinberg_Scaling,MOROZOV_PNAS,Alert2020}.} %\vedcomment{CHECK: can add: ``although the observed scaling depends on the microstructural state and the nature of the imposed or internal forcing''}
%\ju{???,  and the two systems exhibit similar universal scaling properties \cite{Steinberg_Scaling,MOROZOV_PNAS,Alert2020}.???}
%\bluestrike{Under certain conditions, both describe ensembles of rodlike particles that act as contractile force dipoles.} 
A key difference, however, is that active nematics are 
%\bluestrike{turbulence can be} 
driven by internal activity, whereas polymeric %\bluestrike{turbulence}
systems require an imposed flow. 
%\su{Previous studies have hinted at similarities between inertialess turbulence in active matter and polymer solutions \cite{ACTIVE_TURB_ANNUAL_REV,MOROZOV_PNAS,Alert2020,Dubief_Coherent}. \citet{Puggioni} recently demonstrated numerically that chaotic flows in dilute suspensions of rodlike polymers display defect dynamics similar to active nematic turbulence.}

The classical kinetic theories of Doi and Edwards \cite{doi1988theory} describe the orientational dynamics of nematics and rod-like polymers, and their coupling to the surrounding flow, using conformation tensors. More recently, \citet{Puggioni} demonstrated numerically that chaotic flows in dilute suspensions of rigid rod-like polymers---a fundamentally passive system---are accompanied by the emergence and interaction of topological defects in the polymer director field, drawing a qualitative parallel with active nematic turbulence. However, this model does not account for polymer stretching and therefore lacks the deformability that is essential to elastic turbulence. Other papers have also highlighted elastic and active turbulence as different examples of chaotic flows at low Reynolds numbers \cite{ACTIVE_TURB_ANNUAL_REV,MOROZOV_PNAS,Alert2020,Dubief_Coherent} but 
%\redstrike{ Nevertheless, these connections have remained largely phenomenological, and} 
a direct mechanistic, continuum-level correspondence between active nematics and polymeric fluids is still lacking.

%Previous studies have indeed hinted at the shared similarities between inertialess turbulence in active matter and polymer solutions \cite{ACTIVE_TURB_ANNUAL_REV,MOROZOV_PNAS,Alert2020,Dubief_Coherent}. Most notably, Puggioni \& Musacchio \cite{Puggioni} recently demonstrated numerically that chaotic flows in dilute suspensions of rigid rod-like polymers---a fundamentally passive system---are accompanied by the emergence and interaction of topological defects in the polymer director field, drawing a qualitative parallel with active nematic turbulence. However, these connections have remained largely phenomenological, and a direct mechanistic and continuum-level correspondence between the two systems is still lacking. 
%\ju{old versions commented out, reordered here - my aim was to highlight the important things we do more, so I have moved the more detailed things to the summary of the paper, and included what we do but not the results which I thought were too hard at this stage - but we will need to discuss what is best}

In what follows, we investigate the correspondence between the continuum descriptions of polymeric fluids and active nematics. We demonstrate a direct theoretical mapping from the constitutive equations of dilute polymer suspensions  to the analogous description for active nematics, which we show corresponds to the limit of constant polymer extension.
We identify dimensionless variables that relate the physics of active nematics, deformable active nematics, and dilute polymer suspensions, thereby connecting their hydrodynamics and instabilities, and reveal an activity-driven flow-suppression transition. The correspondence reveals nematic signatures exhibited by stretched polymers, and we highlight the link between topological defect pairs in the nematic ordering of stretched polymers and the distinctive arrowhead coherent structures observed  in polymeric fluids \cite{Page_PRL,Morozov_Coherent,Lellep_Linkmann_Morozov_2023,Lewy_Kerswell_2025} when subjected to Kolmogorov flow (Fig.~\ref{introfig}A,C).
Moreover, the mapping motivates a hydrodynamic description of active nematics that allows changes in nematogen shape, which is important
% \cyanstrike{for many epithelial cells.} 
in modeling biological systems such as epithelial tissues.

%In this continuum picture, polymer stretching can be considered to generate an effective activity coefficient that varies in space and time. The correspondence reveals nematic signatures exhibited by stretched polymers, and we highlight the link between topological defect pairs in the nematic ordering of stretched polymers and the distinctive arrowhead coherent structures observed for elastic turbulence \cite{Page_PRL,Morozov_Coherent,Lellep_Linkmann_Morozov_2023,Lewy_Kerswell_2025} \su{when subjected to Kolmogorov flow, \vedcomment{see Fig.~\ref{introfig}C}}.

The paper is organized as follows: Section~\ref{theory} of the paper introduces the equations of motion for a dilute polymeric fluid, and for active nematics, and describes the mapping between the polymers and %\cyanstrike{driven,} 
contractile active nematics. Simulation details are summarized in Section~\ref{sec:Simulation details}. In Section~\ref{results}, we present numerical solutions of the equations of motion to illustrate the connection between the two systems. Our analysis follows a three-step approach. We first present a phase diagram illustrating the diverse behavior of polymeric systems under Kolmogorov (sinusoidal) forcing. We then keep the polymer extension fixed, corresponding to the active nematic limit, and illustrate the 1D instabilities exhibited by the system. We next allow the polymer extension to vary illustrating the onset of 2D instabilities. Then, in Section~\ref{Sec:gradient}, we  compare active nematics with 
% \bluestrike{in space, but not in time, which corresponds to allowing}
spatial variations in  activity  to
% \bluestrike{. Finally we compare to} 
solutions of the full polymeric equations, where the polymer extension evolves dynamically in both space and time. Section~\ref{discussion} summarizes and discusses our results.

\section{Mapping the equations of motion of polymeric fluids to active nematic fluids.}\label{theory}
% ALTERNATIVE:

\subsection{\textbf{Equations of motion for a dilute suspension of polymers}}

To investigate their microstructure under flows, polymers are canonically represented by a 
%\cyanstrike{space-time dependent, symmetric conformation tensor} 
spatially and temporally varying field described by the conformation tensor, $\mathsfbi{C}$, 
where $\operatorname{tr}(\mathsfbi{C})$ is a measure of the polymer extension.
%\ju{may not be needed} This can be written in terms of its eigenvalues $\lambda_1,\lambda_2$ and largest eigenvector $\bm{n}$ as
%\begin{equation}
%\mathsfbi{C}= (\lambda_1-\lambda_2)\left(\bm{n}\bm{n}-\mathsfbi{I}/2\right) +(\lambda_1+\lambda_2) \mathsfbi{I}/2,
%\label{decomposition}
%\end{equation}
%where $(\lambda_1+\lambda_2)$ is a measure of polymer extension, \ju{and $(\lambda_1-\lambda_2)$ is a measure of polymer alignment along  $\bm{n}$  (see supplementary material)}. 
The equation of motion of the conformation tensor is 
\begin{equation}\label{eq:2}
\frac{D\mathsfbi{C}}{D t} = \mathsfbi{S}^{\mathsfbi{C}}  -\frac{1}{\tau_p}\left(\mathsfbi{C}-\mathsfbi{I}\right)+\varepsilon\bnabla^2\mathsfbi{C}.
\end{equation}
On the lhs of Eq.~\eqref{eq:2}, $D\bullet/D t = {\partial_t \bullet} + \mathbf{u} \bcdot \bnabla  \bullet$ 
% \cyanstrike{is} 
denotes the material derivative. The first term on the rhs is the generalized convected %\ved{[Note: might be more appropriate to generally refer to this as the convected derivative term, as it can be lower-or-upper-convected depending on the sign of $\lambda$]} 
derivative
 \begin{equation}   
  \mathsfbi{S}^{\mathsfbi{C}}=  \mathsfbi{C}\bcdot\left(\bnabla\mathbf{u}\right)+\left(\bnabla\mathbf{u}\right)^{\mathsf{T}} \bcdot \mathsfbi{C}+(\lambda-1)(\mathsfbi{C}\bcdot\mathsfbi{E}+\mathsfbi{E}\bcdot\mathsfbi{C}),
\end{equation}
which describes the response of the polymers to velocity gradients where $\mathbf{u}$ is the velocity field, $\mathsfbi{E}$ is the strain rate tensor, and ${\bm{\Omega}}$ %\cyanstrike{(used below)} 
is the vorticity tensor. $\lambda$ is the flow aligning parameter, which is introduced as a material-dependent quantity describing the relative effects of strain rate and vorticity on the microstructure. The second term on the rhs of Eq.~\eqref{eq:2} models relaxation of the polymers to their unstretched state, which corresponds to the unit tensor $\mathsfbi{I}$, on a time scale $\tau_p$. The final term is a diffusion term with coefficient $\varepsilon$ \cite{gupta_vincenzi_2019,Dzanic_PRE,Gupta_Vincenzi_2024}.%\su{this bit can go after the mapping I think.}
%\ju{I've put f=1}

The velocity field obeys the incompressible Navier-Stokes equations,
\begin{equation}\label{eq:1}
    \textcolor{black}{\bnabla \bcdot \mathbf{u} = 0,} ~~ \rho\frac{D\mathbf{u}}{D t}=-\bnabla P + \mu_s \bnabla^2\mathbf{u}+\bnabla \bcdot \bm{\sigma}_p + \rho F\mathbf{\hat{F}},
\end{equation}
where $\rho$, $P$, $\mu_s$ are respectively the density, pressure and solvent's dynamic viscosity, and $F\mathbf{\hat{F}}$ is an external driving force density of magnitude $F$ along the unit vector $\mathbf{\hat{F}}$. The additional stress exerted on the fluid as the polymers relax towards equilibrium is 
\begin{equation} \label{polymerstress}
\bm{\sigma}_p=\frac{\mu_p}{\tau_p}\left(\mathsfbi{C}-\mathsfbi{I}\right),
\end{equation}
 where $\mu_p$ is the dynamic viscosity of the polymers \cite{HINCH2021104668,Stone_Deriv}. 
 %\ju{is this form a sensible guess or exact for dumbbells?}

\subsection{\textbf{Equations of motion for active nematics}}
In a similar approach the microstructure of an active nematic is encoded by a symmetric traceless tensor, 
\begin{equation}
\mathsfbi{Q}= 2S \left(\bm{n}\bm{n}  -\mathsfbi{I}/2\right) 
\end{equation}
in two-dimensions, where $S$ is the magnitude of the nematic order along a director $\bm{n}$ which is the  eigenvector corresponding to the largest eigenvalue of $ \mathsfbi{Q}$.
 The equation of motion for $\mathsfbi{Q}$ is
\begin{equation}
    \frac{D\mathsfbi{Q}}{D t} = \mathsfbi{S}^{\mathsfbi{Q}} + \Gamma \mathsfbi{H},
    \label{Q-equation}
\end{equation} 
with a generalized co-rotational term 
\begin{multline}\label{Q-rotation}
\mathsfbi{S}^{\mathsfbi{Q}}=(\lambda \mathsfbi{E} + {\bm{\Omega}})\cdot (\mathsfbi{Q}+\mathsfbi{I}/2)+(\mathsfbi{Q}+\mathsfbi{I}/2)\cdot(\lambda \mathsfbi{E} - {\bm{\Omega}})
   \\ -2\lambda(\mathsfbi{Q}+\mathsfbi{I}/2)(\mathsfbi{Q}:\mathsfbi{E}).
   \end{multline}
The second term on the rhs of Eq.~\eqref{Q-equation} describes the relaxation of the nematogens to the minimum of a free energy at a rate $\Gamma$. We choose a free energy \cite{deGennes1995}
\begin{equation}\label{eq:freeenergyAK}
    {\cal F} = \frac{A}{2}\mathsfbi{Q}^2 %+   \frac{B}{3}\mathsfbi{Q}^3 + \frac{C}{4}\mathsfbi{Q}^4 
    + \frac{K}{2}(\bnabla\mathsfbi{Q})^2,
\end{equation} 
where $A$ is a material coefficient, which is taken to be positive to ensure that the system is in a paranematic (isotropic) state with $S=0$ in the absence of activity and flows \cite{Santhosh2020}.
$K$ is the elastic constant, assuming the usual single elastic constant approximation. The molecular field is then
\begin{equation}\label{free}
    \mathsfbi{H} = -\frac{\delta {\cal F}}{\delta \mathsfbi{Q}} + \frac{\mathsfbi{I}}{2} \operatorname{tr}\left(\frac{\delta {\cal F}}{\delta \mathsfbi{Q}}\right)  = -A\mathsfbi{Q} + K\bnabla^2\mathsfbi{Q}.
 %   -B\mathsfbi{Q}^2 -C\mathsfbi{Q}^3 
\end{equation} 
%Substituting Eq.~\eqref{free} into Eq.~\eqref{Q-equation} and writing in terms of dimensionless variables leads to
%\begin{equation}
%   \frac{D\mathsfbi{Q}}{D t} = \mathsfbi{S}_{\mathsfbi{Q}} -\frac{1}{(Wi)^\mathsfbi{Q}}\mathsfbi{Q}+\frac{1}{(Pe)^\mathsfbi{Q}} \bnabla^2\mathsfbi{Q},
%\end{equation}
%where 
%\begin{equation}
%\frac{1}{(Wi)^\mathsfbi{Q}}=\frac{\Gamma A L}{U_0}, \;\;\;\; (Pe)^\mathsfbi{Q}=\frac{U_0 L}{\Gamma K}.
%\end{equation}

The velocity field for active nematics is described by Eq.~\eqref{eq:1}, but with the polymer stress replaced by the sum of an active stress,  $\bm{\sigma}_{active}$, created by the active dipolar forces produced by individual particles \cite{LSA_Ramaswamy}, together with a small elastic backflow term %$\bm{\sigma}_{elastic}$, 
associated with the relaxation of the order parameter to the free energy minimum, which we will neglect. The active stress is
\begin{equation}\label{active stress}
\bm{\sigma}_{active}=-\zeta \mathsfbi{Q},   %\;\;\;\;\;\;\;\; \bm{\sigma}_{elastic}= A \lambda\,(1-4S^{2})\,\mathsfbi{Q}.
\end{equation}
 where $\zeta$ is the activity coefficient which quantifies the strength of the activity where  
 % \cyanstrike{$\zeta<0$ corresponds to contractile activity, which will primarily concern us here, where the dipolar forces act inwards along a particle’s axis of elongation; $\zeta>0$ represents extensile activity, where the forces act outwards along the long axis.} 
$\zeta<0$ ($\zeta > 0$) corresponds to contractile (extensile) activity \cite{Doostmohammadi18}.

\subsection{\textbf{Comparing polymer and active nematic hydrodynamics}}\label{SS:Mapping}

The key difference between the conformation tensors $\mathsfbi{C}$ and $\mathsfbi{Q}$ is that the latter is traceless. Therefore we construct a normalized, traceless tensor, $\mathsfbi{C^*} = (\mathsfbi{C} - \operatorname{tr}(\mathsfbi{C})\mathsfbi{I}/2)/\operatorname{tr}(\mathsfbi{C})$, and decompose the polymer tensor into a traceless component, $\operatorname{tr}(\mathsfbi{C}) \mathsfbi{C^*}$, and a diagonal component, $\operatorname{tr}(\mathsfbi{C})\mathsfbi{I}/2$ by writing
\begin{equation}
\mathsfbi{C}=\operatorname{tr}(\mathsfbi{C})(\mathsfbi{C^*}+\mathsfbi{I}/2).
\label{decomp}
\end{equation}
Substituting this expression into Eq.~\eqref{eq:2}, and after some algebra, which is detailed in the Supporting Information, we derive the following governing equations for $\mathsfbi{C^*}$ and $\operatorname{tr}(\mathsfbi{C})$:
\begin{equation}\label{Cstar-equation}
  \frac{D \mathsfbi{C^*}}{D t}
    = \mathsfbi{S}^{\mathsfbi{C^*}} - \frac{2}{\operatorname{tr}(\mathsfbi{C})\tau_p}(\mathsfbi{C^*})    
    + \varepsilon\bnabla^2\mathsfbi{C^*} +\frac{2\varepsilon}{\operatorname{tr}(\mathsfbi{C})}\bnabla\operatorname{tr}(\mathsfbi{C})\cdot\bnabla\mathsfbi{C^*},
\end{equation}
where
\begin{multline}\label{Cstar-rotation}
    \mathsfbi{S}^{\mathsfbi{C^*}}=(\lambda \mathsfbi{E} + {\bm{\Omega}})\cdot (\mathsfbi{C^*}+\mathsfbi{I}/2)+(\mathsfbi{C^*}+\mathsfbi{I}/2)\cdot(\lambda \mathsfbi{E} - {\bm{\Omega}})
   \\ -2\lambda(\mathsfbi{C^*}+\mathsfbi{I}/2)(\mathsfbi{C^*}:\mathsfbi{E}),
\end{multline}
\begin{multline} \label{Eq:4}
    \frac{D~\operatorname{tr}(\mathsfbi{C})}{D t} =  2\lambda\operatorname{tr}(\mathsfbi{C})\left(\mathsfbi{C^*}:\mathsfbi{E} \right)
\\    -\frac{1}{\tau_p} \left( \operatorname{tr}(\mathsfbi{C})-2 \right) 
    +\varepsilon\bnabla^2\left(\operatorname{tr}(\mathsfbi{C})\right).
\end{multline} 

Comparing the microstructural evolution equations for ${\mathsfbi{C^*}}$, Eq.~\eqref{Cstar-equation} and Eq.~\eqref{Cstar-rotation} 
% \cyanstrike{in the rigid rod limit} 
at fixed polymer extension, $\operatorname{tr}(\mathsfbi{C}) = $ constant, with those for the nematic tensor, $\mathsfbi{Q}$, Eq.~\eqref{Q-equation}, Eq.~\eqref{Q-rotation} and Eq.~\eqref{free}, shows that they are identical if
\begin{eqnarray}
 %\{-\zeta+A\lambda\,(1-4S^{2})\}\mathsfbi{Q} & \Leftrightarrow & \operatorname{tr}(\mathsfbi{C})\frac{\mu_p}{\tau_p}\left(\mathsfbi{C^*}+\frac{\mathsfbi{I}}{2}\right),\label{stressmap} \\
  -\Gamma A\mathsfbi{Q} %-B\mathsfbi{Q}^2-C\mathsfbi{Q}^3 
  & \Leftrightarrow  &  - \frac{2}{\operatorname{tr}(\mathsfbi{C})\tau_p}(\mathsfbi{C^*}),  \label{freemap}\\
  \Gamma K\bnabla^2\mathsfbi{Q} & \Leftrightarrow & \varepsilon\bnabla^2\mathsfbi{C^*}. \label{diffusionmap}
\end{eqnarray}
Similarly, the equations of motion for the velocity fields of the two tensors map onto each other if the deviatoric part of the polymeric and active stresses, Eq.~\eqref{polymerstress} and Eq.~\eqref{active stress}, are related by
 \begin{eqnarray}
 -\zeta\mathsfbi{Q} & \Leftrightarrow & \frac{\mu_p}{\tau_p}\left(\operatorname{tr}(\mathsfbi{C})\mathsfbi{C^*}
 %\bluestrike{+\frac{(\operatorname{tr}(\mathsfbi{C})-2)}{2}{\mathsfbi{I}}}
 \right).\label{stressmap} 
\end{eqnarray}
Thus the polymer equations map onto those of active nematics if Eq.~\eqref{Eq:4} can be neglected, i.e. if $\operatorname{tr}(\mathsfbi{C})$ is held constant. While this reflects the rigid-limit correspondence between polymers and liquid crystals, our parameter mapping reveals additional insights into the underlying dynamics of the two systems:

%\bluestrike{Before presenting numerical results demonstrating the connection between elastic and active low Reynolds number turbulence, we comment on the parameter mapping:} 
\begin{itemize}
    \item Eq.~\eqref{stressmap} compares the active and polymeric stresses which will lead to the same forces if the activity coefficient, $\zeta=-\frac{\mu_p}{\tau_p}\operatorname{tr(\mathsfbi{C})}$.  Importantly the correspondence holds for $\zeta < 0$, corresponding to contractile activity that exerts an active stress returning the nematic system  to an isotopic, configuration, just as the elastic stress returns the polymers to an isotropic, unstretched state. 
    %As expected, higher $\operatorname{tr(\mathsfbi{C})}$, i.e.~more elongated polymers, and shorter relaxation times, ${\tau_p}$, correspond to higher activities.}  
%\bluestrike{\rv{Eq.~\eqref{stressmap} establishes the correspondence between the active and polymeric stresses, which generate equivalent forces when the activity coefficient is $\zeta=-\frac{\mu_p}{\tau_p}\operatorname{tr}(\mathsfbi{C})$. Importantly, this correspondence requires $\zeta<0$, corresponding to contractile activity. The resulting active stress drives the nematic towards an isotropic configuration, analogous to the elastic stress that returns the polymers towards an isotropic, unstretched state.}} %As expected, greater polymer extension, quantified by $\operatorname{tr}(\mathsfbi{C})$, and shorter relaxation times, $\tau_p$, correspond to larger activity magnitudes.} 
The equivalent activity coefficient $\zeta$ is therefore determined by both the  polymer extension, $\operatorname{tr}(\mathsfbi C)$, and the strength of the polymeric elastic response, captured by $\mu_p/\tau_p$.
% Note: could also define it as elastic modulus...

%{\su{Recalling that gradients of the stress control flow dynamics, in Section~\ref{results} we investigate this correspondence between the two systems in more detail.}}

\item 
% \cyanstrike{Eq.~\eqref{freemap} is the mapping of the molecular field, which encodes relaxation to equilibrium. The mapping is physically sensible  for a positive quadratic term in the free energy (Eq.~\eqref{eq:freeenergyAK}), $A > 0$, and hence an equilibrium state where there is no nematic alignment,  $S=0$ corresponds to the isotropic state, namely the state at high temperature. It is known that such an isotropic equilibrium state is stable to contractile activity in the absence of imposed flow \cite{Santhosh2020}, corresponding to the occurrence of polymer dynamics only under flow.} 
Eq.~\eqref{freemap} maps the molecular field, which governs relaxation towards equilibrium. The correspondence is physically consistent for $A>0$ in Eq.~\eqref{eq:freeenergyAK}, for which the equilibrium state is isotropic, with $S=0$, corresponding to the high-temperature phase of a nematic. This isotropic state is stable under contractile activity in the absence of imposed flow \cite{Santhosh2020}, consistent with polymer dynamics arising only when flow is imposed.
%The relaxation to equilibrium is faster for a steeper potential well (larger $A$) in the active nematic case, or more rapid relaxation once the driving is removed (smaller $\tau_p$) for polymers. 

\item 
%\cyanstrike{Finally, Eq.~\eqref{diffusionmap} shows that the diffusion of stretched polymers corresponds to Frank elasticity, which describes deformations of the orientational order in liquid crystals and active nematics. In the polymer literature, $\varepsilon$ is typically chosen as small as possible while maintaining numerical stability on the premise that this limit best represents physically realistic polymer behavior. However its value is known to influence quantitative features of the flow \cite{gupta_vincenzi_2019,Dzanic_PRE,Gupta_Vincenzi_2024}. By contrast, in  active nematic systems, the elastic constant $K$ is an important contributor to the physics: in particular the active length scale, which describes the size of vortices and of nematic domains in the active turbulent state $\sim \sqrt{K/\zeta}$ \cite{Sumesh_Euro,Thampi2014}.}
Finally, Eq.~\eqref{diffusionmap} shows that polymer diffusion maps onto Frank elasticity, which penalizes spatial distortions of orientational order in liquid crystals and active nematics \cite{deGennes1995}. In simulations of polymeric flows, $\varepsilon$ is typically chosen to be as small as numerical stability permits, based on the assumption that the small-diffusivity limit most closely represents physical polymer behavior. Nevertheless, $\varepsilon$ is known to influence quantitative features of the flow \cite{gupta_vincenzi_2019,Dzanic_PRE,Gupta_Vincenzi_2024}. By contrast, in active nematics, the corresponding elastic constant $K$ plays a central physical role. In particular, together with the activity, it sets the active length scale, $\ell_a \sim \sqrt{K/|\zeta|}$, which characterizes the typical size of vortices and nematic domains in the active turbulent state \cite{Sumesh_Euro,Thampi2014}.
%\su{The correspondence between $\varepsilon$ and $K$ therefore suggests that choosing $\varepsilon > 0$ may not be simply `an unphysical numerical regularization', but rather a physically meaningful modeling choice in polymeric systems.} \ju{don't entirely agree, would have to have polymers that are keen on aligning - maybe we can mention elastomers}
\end{itemize}

\begin{figure*}[t]
	\centering
    \includegraphics[width=\textwidth]{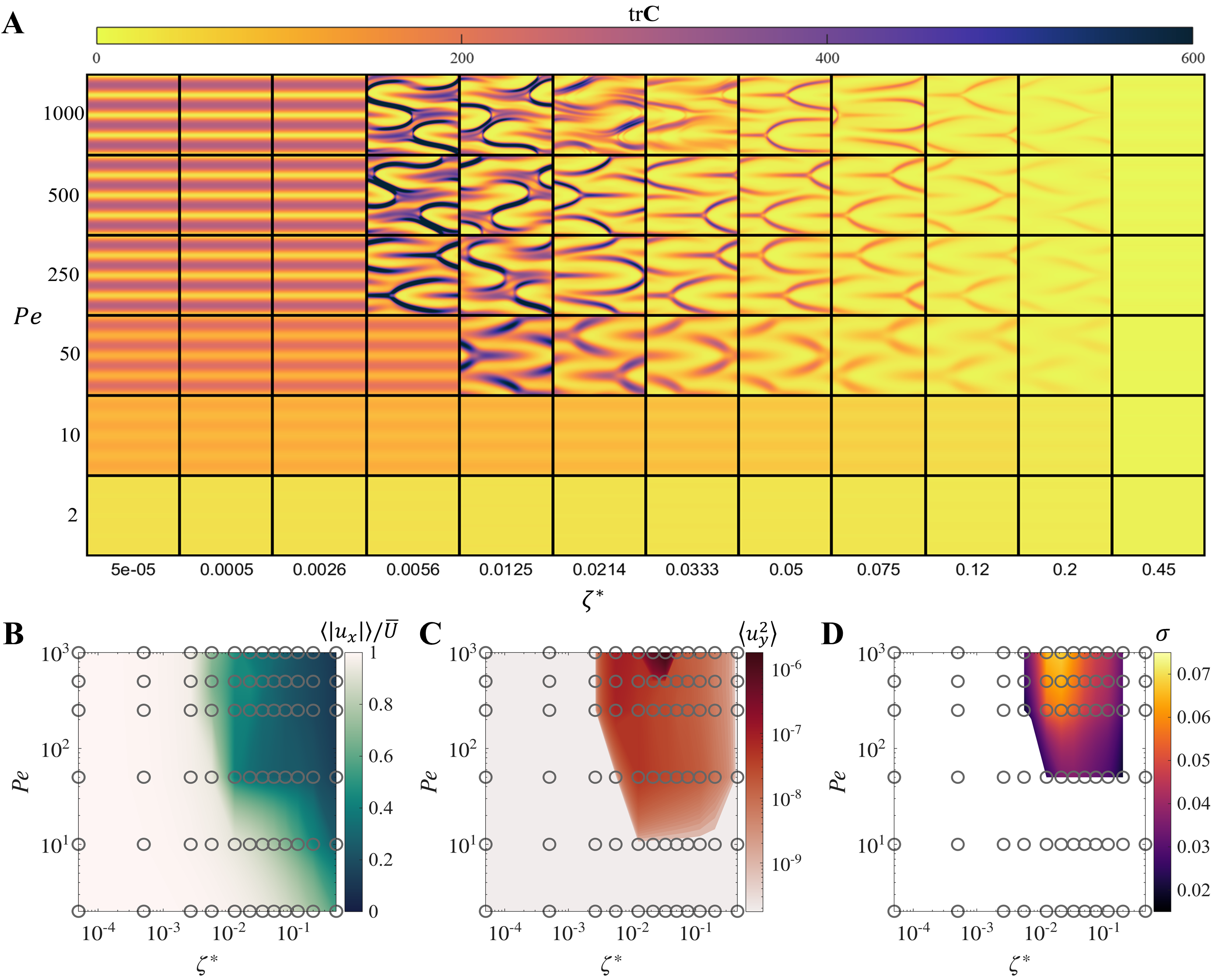}
	\caption{\textbf{Phase diagram in the P\'eclet  number $Pe$ -- dimensionless activity coefficient $\zeta^{\ast}$  plane illustrating the long time behavior of the polymer flows when subjected to Kolmogorov forcing.} (\textbf{A}) Representative $\operatorname{tr(\mathsfbi{C})}$ fields at $Wi = 20$; the scale bar is shown above the phase diagram. (\textbf{B}) The streamwise velocity magnitude,  normalised by the mean Kolmogorov speed $\langle|u_x|\rangle/\bar{U}$, where $\bar{U} = \frac{1}{2\pi}\int_0^{2\pi} U|\cos y|\,dy = \frac{2U}{\pi}$. (\textbf{C}) The mean-square velocity $\langle u_y^2 \rangle$ of the $y$-component of the flow. (\textbf{D}) The initial growth rate of the transverse instability, defined by $\sigma = \frac{1}{2}\frac{d}{dt}\ln\langle u_y^2\rangle$, in the same
    $Pe$--$\zeta^{\ast}$ plane.}
	\label{phase_diagram}
\end{figure*}

\begin{figure*}[t]
	\centering
	\includegraphics[width=1.0\textwidth]{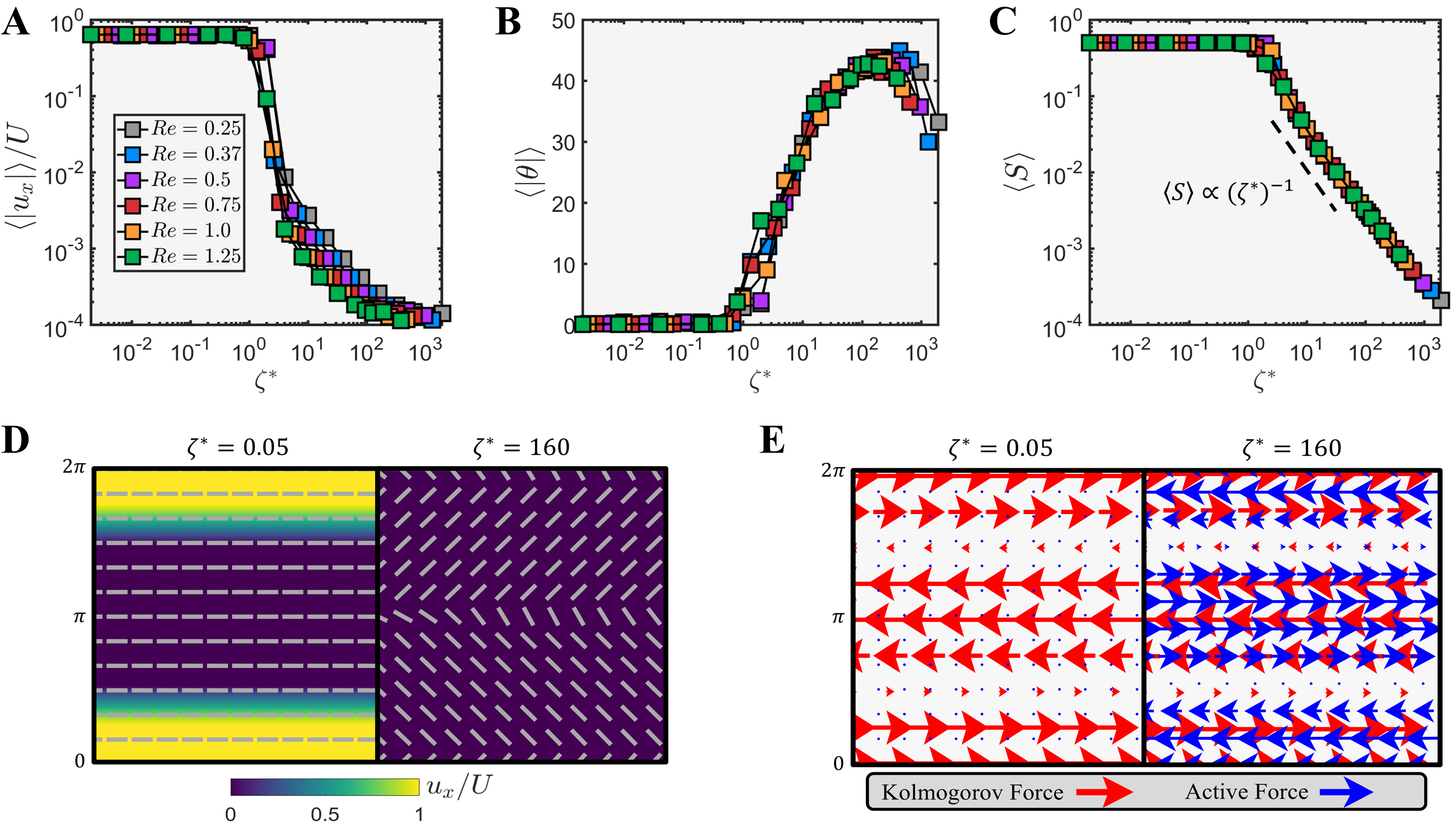}%width=0.95\textwidth
	\caption{\textbf{Spontaneous flow instability in inextensible polymers (active nematics) with a %\bluestrike{constant} 
    contractile activity coefficient}. Simulations are performed across the $Re -\zeta^*$ parameter space for the spatially averaged (\textbf{A}) normalized streamwise velocity magnitude $\langle\lvert u_x \lvert \rangle/U$,  (\textbf{B}) angle between the director and the $x$-axis, $\langle\lvert\theta\rvert\rangle$, and (\textbf{C}) order parameter magnitude $\langle S\rangle$.  A rapid change in behavior occurs at $\zeta^*\approx10^0$, marking the point at which contractile active stresses begin to impact the dynamics. (\textbf{D}) This behavior is further illustrated in the streamwise-velocity contour plots comparing the low ($\zeta^*=0.05$) and high ($\zeta^*=160$) activity regimes. At low activity, the system recovers the passive Kolmogorov forcing laminar solution, with a well-ordered director field (white bars) aligned in the streamwise direction. As the activity is increased, the flow is strongly reduced and the director field exhibits pronounced splay. (\textbf{E}) Corresponding force fields:  contractile active forces (blue) and Kolmogorov forcing (red). The flow-suppressed state at $\zeta^{\ast} = 160$ corresponds to two forces
    % \bluestrike{This jammed state corresponds to  contractile active forces (blue) that are} 
    of comparable magnitude 
    % \bluestrike{to the imposed Kolmogorov forcing (red),}
    but opposite in direction, resulting in a near-complete cancellation of the mean flow.}
	\label{fig2}
\end{figure*}

%%
%\ju{??It is expected that the diffusion constant in small in polymeric systems, and that increasing its value to render simulations more stable, gives quantitative changes in the details of the turbulent state. It is interesting that increasing $\varepsilon$ leads to increased finite size effects in the energy spectrum. In active turbulence the length scale of vortices is $\sqrt{K/{\zeta}}$ and so increasing $K$ implies larger vortices and more finite size effects??} 

%\ju{To summarise, if the trace of the polymer conformation tensor, i.e. the polymer shape, is fixed, the equations of motion for elastic turbulence in polymeric systems are identical to those of contractile active nematics for which the magnitude of the nematic order is zero in the absence of activity.	In the polymeric system the primary (only?) effect of allowing the trace to vary is to allow a variation in the activity in space and time.}
\subsection{\textbf{Non-dimensionalizing the equations of motion}}
Our numerical results are obtained by solving~Eq.~\eqref{eq:2} for the conformation tensor $\mathsfbi{C}$ 
%Eq.~\eqref{Eq:4} for $\operatorname{tr}(\mathsfbi{C})$, which encodes the polymer extension, 
and~Eq.~\eqref{eq:1} for the velocity field ${\bf u}$. 
It is helpful to non-dimensionalize these equations  as 
\begin{equation}
   Re \frac{D\mathbf{u}}{D t}=-\bnabla P + \bnabla^2 \mathbf{u}+\frac{\mu_p}{\mu_s} \frac{1}{Wi}\bnabla \bcdot \left(\mathsfbi{C}-\mathsfbi{I}\right) + Re_F\mathbf{\hat{F}},
   \label{nondimu}
\end{equation}
\begin{equation}
\frac{D\mathsfbi{C}}{D t} = \mathsfbi{S}^{\mathsfbi{C}}  -\frac{1}{Wi}\left(\mathsfbi{C}-\mathsfbi{I}\right)+\frac{1}{Pe}\bnabla^2\mathsfbi{C},
\label{nondimC}
\end{equation}
where we do not explicitly distinguish the dimensionless fields to avoid cumbersome notation. The dimensionless control parameters, the Reynolds number $Re$, the flow Reynolds number $Re_F$,  the Weissenberg number $Wi$ and the P\'eclet number $Pe$ are, respectively, defined as
\begin{equation}
 Re=\frac{\rho U L}{\mu_s}, \;\; Re_F=\frac{\rho F L^2}{\mu_s U}, \;\; 
 \frac{1}{Wi}=\frac{L}{\tau_p U},\;\; 
 Pe=\frac{U L}{\varepsilon},
\end{equation}
where $U$ and $L$ are characteristic velocity and length scales.
 Note that in the active nematic limit of constant $\operatorname{tr}(\mathsfbi{C})$, Eq.~\eqref{stressmap} can be used to define a dimensionless activity $\zeta^\ast$ as
\begin{equation}
  \frac{\mu_p}{\mu_s}\frac{1}{Wi}=\frac{\zeta L}{\mu_s U}  \equiv \zeta^\ast.
\end{equation}
Hence, in terms of non-dimensional variables, the viscosity ratio normalized by the Weissenberg number acts as the system’s activity parameter in polymeric flows. 

% \section*{\redstrike{Results}}
% \label{results}
% \subsection{\redstrike{\textbf{Simulation details}}}

\section{Simulation details}
\label{sec:Simulation details}

We solve the equations of motion, driving the system by imposing
%parallel periodic shear flow using 
 two-dimensional Kolmogorov flow with input energy along ${\mathbf{\hat x}}$,
%(Materials and Methods)
%We consider a two-dimensional system and, to input energy, we apply a Kolmogorov flow along ${\mathbf{\hat x}}$,
\begin{equation}
    F(y) = F_{{KF}} \cos({y}),
    \label{Kolmogorov}
\end{equation}
where $F_{KF}$ is the forcing amplitude and we have chosen the unit of length so that the forcing wavelength is $2\pi$. As illustrated in Fig.~\ref{introfig}C, the domain is of size $\left[0, 2\pi n\right]\times\left[0, 2\pi n\right]$ with fully periodic boundary conditions, where the
integer $n$ defines the number of wavelengths applied to the system. By using the amplitude of the fixed-point Newtonian laminar velocity (i.e., $\mu_p=0$) $U=\rho F_{KF}L^2/\mu_s$, we determine the turnover time $T=\mu_s/\rho F_{KF}L$. 

We take the flow to be in the low-inertia regime, and, as such, keep the Reynolds number $Re<\sqrt{2}$ \cite{Boffetta_PRE_2005,Gotoh1984InstabilityOA}. %\bluestrike{The diffusion coefficient is fixed by setting $Pe\approx2.5\cdot10^{2}$, and the} 
% \redstrike{The activity is controlled by varying the polymer viscosity $\mu_p$.} 
% \redstrike{For results presented below, $Pe\approx2.5\cdot10^{2}$ except for Fig.~\ref{phase_diagram}, where $Pe$ is varied by changing the polymer diffusivity $\varepsilon$.}
%Moreover, in \textbf{Theory and Model Comparison}, we have demonstrated that much like the order tensor $\mathsfbi{Q}$ in active nematics, the conformation tensor $\mathsfbi{C}$ is governed by a convected time derivative, the nature of which depends on $\lambda$. 
The flow aligning parameter is chosen to be $\lambda=1$, corresponding to the Oldroyd-B model 
commonly used to describe the evolution of polymers \cite{HINCH2021104668,Alves_Rev}. We solve the governing equations using a hybrid lattice Boltzmann–finite-difference method \cite{DZANIC2022105280}, as detailed in \textbf{Materials and Methods}.  

\section{Results}
\label{results}

Taking the dimensionless activity $\zeta^*$ and Péclet number $Pe$ as control parameters, we first construct a phase diagram %\redstrike{below}
to identify the regimes of polymeric and active-nematic behavior. In the subsequent sections, we examine the different %\redstrike{various} 
regions of the phase diagram to discover common physics, thus establishing a bridge between 
%\redstrike{these systems} 
active nematics, deformable active nematics and polymers.

\subsection{\textbf{Phase diagram}}
The long-time behavior of the continuum equations of motion in the space of P\'eclet  number $Pe$ and the activity parameter $\zeta^\ast$ is shown in Fig.~\ref{phase_diagram}A. The dimensionless activity is changed by varying the polymer viscosity $\mu_p$ at a fixed Weissenberg number, $Wi=20$. The colour scale denotes $\operatorname{tr(\mathsfbi{C})}$ which provides an indication of how the polymer extension varies within the simulation domain in each snapshot. Large P\'eclet number (small polymer diffusivity) corresponds to parameters typically used in simulations of polymer elastic turbulence \cite{Lewy_Kerswell_2025, MOROZOV_PNAS, gupta_vincenzi_2019}. In contrast, small P\'eclet number (large polymer diffusivity) suppresses spatial variations in $\mathsfbi{C}$, yielding a more uniformly stretched polymer configuration characteristic of active nematic-like behavior.

Figs.~~\ref{phase_diagram}B--D show, respectively, the average speed along $x$, normalized by the mean Kolmogorov speed, the mean-square velocity of the $y$-component of the flow,
% average kinetic energy density associated with the velocity along $y$, 
and the initial growth rate of  this transverse velocity. 
%\ju{check exact definitions}. 
The plots in Figs.~\ref{phase_diagram}A--D suggest that the parameter space can be usefully divided into three regions: a characteristic ``U-shaped" area 
% \redstrike{region} 
in the center of the phase diagram, and two regions on either side.  

% \redstrike{In the region} 
To the left of the U-shaped region, activity $\zeta^\ast$ is small and, as expected, the active stress has little effect on the imposed Kolmogorov flow. As $Pe$ is increased, the polymers are more readily deformed by the shear gradients of the Kolmogorov flow leading to 
% \redstrike{an} 
increased variation in $\operatorname{Tr(\mathsfbi{C})}$ along $y$. 

Surprisingly, as the activity is increased at fixed $Pe$, the flow velocity in the $x$-direction decreases for all $Pe$ (Fig.~\ref{phase_diagram}B).  At very large activity $\zeta^\ast$, including to the right of the U-shaped region, the flow velocity along $x$ nearly vanishes. The decreasing flow is accompanied by a reduction in polymer extension with $\operatorname{Tr(\mathsfbi{C})}$ eventually approaching that of undeformed polymers, similar to those in the bottom-left, low-$\zeta^*$ -- low-$Pe$ corner of the phase diagram. Movie~S1 %\jucomment{number from 1?} 
shows that this is a persistent flow-suppressed state.
%\ju{(similar to those in the bottom-left, low-$\zeta^\ast$ --  low-$Pe$, corner of the phase diagram)}.

%Typical snapshots of the late time behavior, for increasing values of the dimensionless activity $\zeta^*$ corresponding to each of these regions, are shown in Fig.~\ref{fig1}E for $x~\varepsilon~ [0,4\pi]$ and $y~\varepsilon~[0,4\pi]$, and corresponding animations are provided in Movies S2, S3, S4, and S5.  (need velocity fields for this figure.

In both these regions there is no flow along the $y$ axis (Figs.~\ref{phase_diagram}C,D), apart from transients that depend on initial conditions. By contrast, in the U-shaped region in the center of the phase diagram, a 2D instability develops, resulting in flows along $y$. Correspondingly, the arrowhead structures characteristic of elastic instabilities  in polymers \cite{Kerswell_Page_2024,Morozov_Coherent,Lewy_Kerswell_2025} appear in the $\operatorname{tr(\mathsfbi{C})}$ field. The relative positions of the arrowheads 
% \redstrike{structures} 
become less regular with increasing $Pe$ as the 2D instabilities transition to fully developed elastic turbulence. The corresponding time evolution is shown in Movies~S2--S5. These distinguish the steady passive state (Movie~S3) from  long-lived coherent arrowhead states (Movies~S2 and S4) and the more strongly time-dependent structures that emerge as the coherent state loses stability (Movie~S5). %\jucomment{renumber movies}
Thus, the progression across the U-shaped region represents not only a change in the instantaneous spatial structure, but also a change from coherent to increasingly irregular dynamics.
% \redstrike{is approached}}. 

Varying the Weissenberg number does not alter the qualitative features of the phase diagram: for example, at Wi=13, a similar phase diagram is obtained, but with a weaker 2D instability, and hence, a smaller region where 2D flow patterns are formed (see Fig.~S2 in the SI).

To investigate and explain this diverse behavior, we consider how the system responds to increasing activity $\zeta^\ast$, first at low $Pe$, where there is no 2D instability, and then at high $Pe$  where the system becomes unstable to 2D flows.

\subsection{\textbf{Increasing the activity at low $Pe$}}
At low $Pe$, variations in the trace of the conformation tensor $\mathsfbi{C}$ are small, and here we expect the system to approximate the active nematic limit of constant $\operatorname{Tr(\mathsfbi{C})}$. To understand the low $Pe$ region of the phase diagram, we therefore consider inextensible polymers (i.e.~active nematics). We observe no transverse instabilities, obtaining a 1D flow field, with no velocity along $y$.

%\ju{need to know the path through the phase diagram in the current Fig 2, hopefully we an say that we are moving across the phase diagram at a small, constant Pe, and then need to state that allowing the Trace to vary makes no difference}

% \subsection*{$\operatorname{Constant Tr(\mathsfbi{C})}$: the active nematic limit}

%Holding $\operatorname{Tr(\mathsfbi{C})}$ constant, to correspond to the active nematic limit, we solve the equations of motion in Kolmogorov flow. In this system,  

Fig.~\ref{fig2}A-C show, respectively, how the spatially averaged magnitude of the scaled velocity, $\langle  \mid \! u_x\! \mid \rangle/U$, the magnitude of the angle between the nematic director and the $x$-axis, $\langle  \mid \! \theta\! \mid \rangle$, 
 and the nematic order parameter, $\langle  S  \rangle$, vary with $\zeta^*$.
 (The nematic order parameter $S$, defined as the largest eigenvalue of $\mathsfbi{C}^*$, is related to the extension and alignment of the polymers.) 
 \begin{figure*}[t]
	\centering
    	\includegraphics[width=1.0\textwidth]{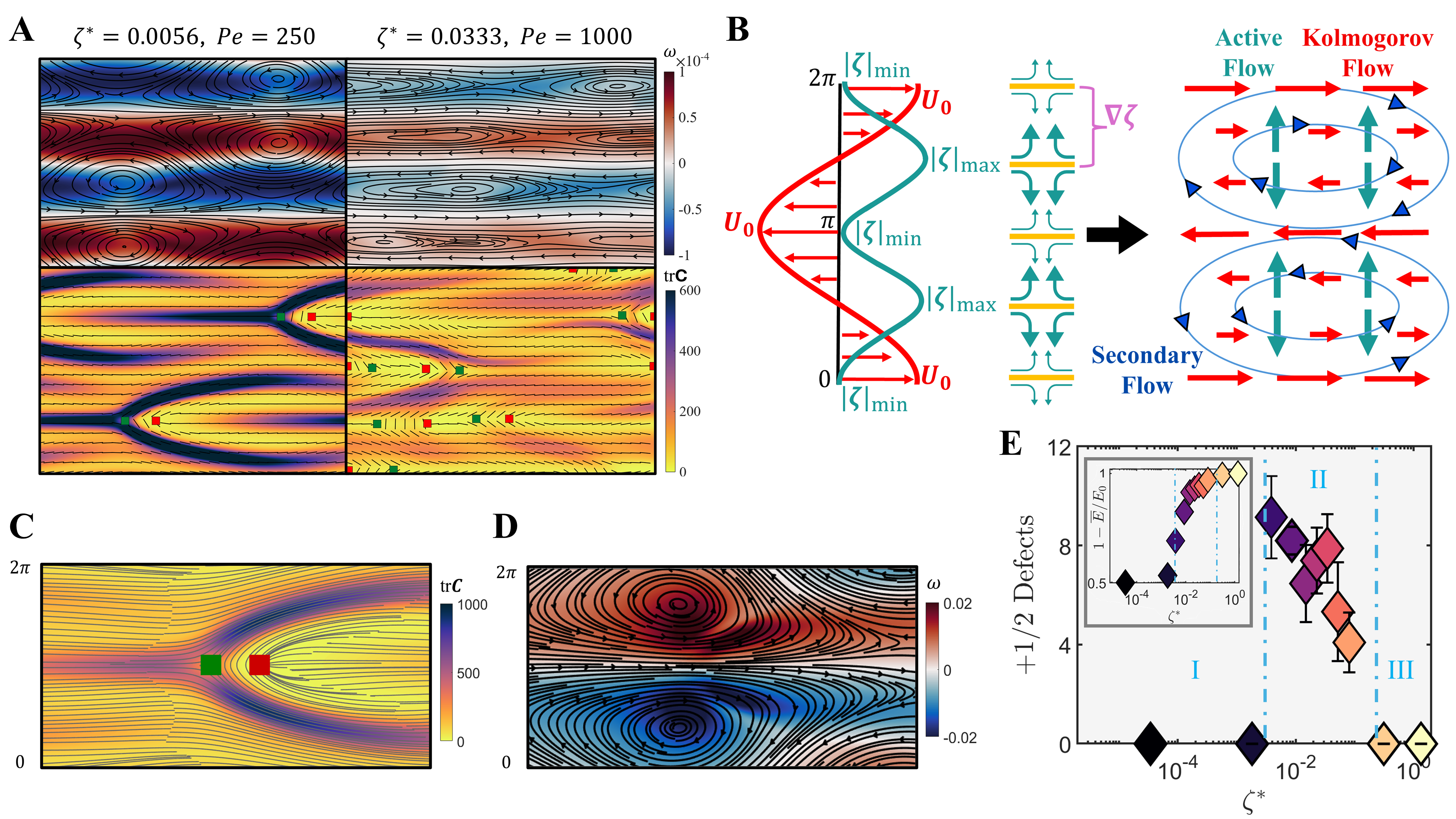}%width=0.95\textwidth
    \caption{
    {\bf Behavior at high Pe: 2D instabilities, arrowheads and topological defects at intermediate activities.} 
    (\textbf{A}) 2D instabilities generating a regular lattice of counter-rotating vortices at $\zeta^{\ast}=0.0056$ and $Pe=250$. At $\zeta^{\ast}=0.0333$ and $Pe=1000$, the vortex lattice becomes unstable and loses its regular structure.
    Top row: vorticity-field contours overlaid with velocity streamlines representing the deviation from the background Kolmogorov flow. Bottom row: corresponding $\operatorname{tr}(\mathsfbi{C})$ field with the director field and $+1/2$ (red) and $-1/2$ (green) defects superimposed.
    (\textbf{B}) Schematic representation of the mechanism through which the 2D instability creates flows in the transverse direction. Each forcing wavelength develops a pair of coherent, counter-rotating vortices. (\textbf{C}) Snapshot of the characteristic arrowhead $\operatorname{tr}(\mathsfbi{C})$ field with the director field and $\pm 1/2$ defects superimposed. (\textbf{D}) Corresponding flow field.  
 %Representative snapshots of the $\operatorname{tr}(\mathsfbi{C})$ field with the director field $\bm{n}$ superimposed for $x=[0,4\pi]$ and $y=[0,4\pi]$ \su{for low and high activities/ corresponding to \textbf{B}}. (\textbf{D}) Snapshot of the \su{association between the} characteristic arrowhead structure \su{and the defect pair} \bluestrike{observed for elastic turbulence} over a single forcing wavelength $y=\left[0,2\pi\right]$. Top: Director field superimposed on top of the $\operatorname{tr}(\mathsfbi{C})$ field, revealing that each arrowhead corresponds to a half-integer defect pair. \su{Bottom: Corresponding flow field} 
 %(\textbf{D}) Time series of the spatially averaged kinetic energy normalized by the kinetic energy of the fixed-point laminar flow, i.e., $E(t)/E_0=\langle \mathbf{u}^2\rangle/U^2$ for different levels of dimensionless activity $\zeta^*$.
%(\textbf{C}) Top: Director field around a comet-like $+1/2$ defect (red marker) and a trefoil-like $-1/2$ defect (green marker). Bottom: Characteristic arrowhead structure in the $\operatorname{tr}\mathsfbi{C}$ field showing the associated topological defects.
 (\textbf{E}) Number of $+1/2$ defects for different $\zeta^*$ in the statistical steady-state. The inset shows the temporally averaged flow resistance $1-\overline{E}/E_0$ for the same values of $\zeta^*$. In both plots, we observe a
 %\bluestrike{non-trivial} 
 sequence of transitions with increasing $\zeta^*$: I. passive Kolmogorov flow with no defects, II. a defect-rich regime accompanied by increasing flow resistance, III. a defect-free, high-resistance, flow-arrested state, the three regions observed in the phase diagram in Fig.~\ref{phase_diagram}.}
	\label{highPe}
\end{figure*}

\begin{figure*}[t]
	\centering
	\includegraphics[width=1.0\textwidth]{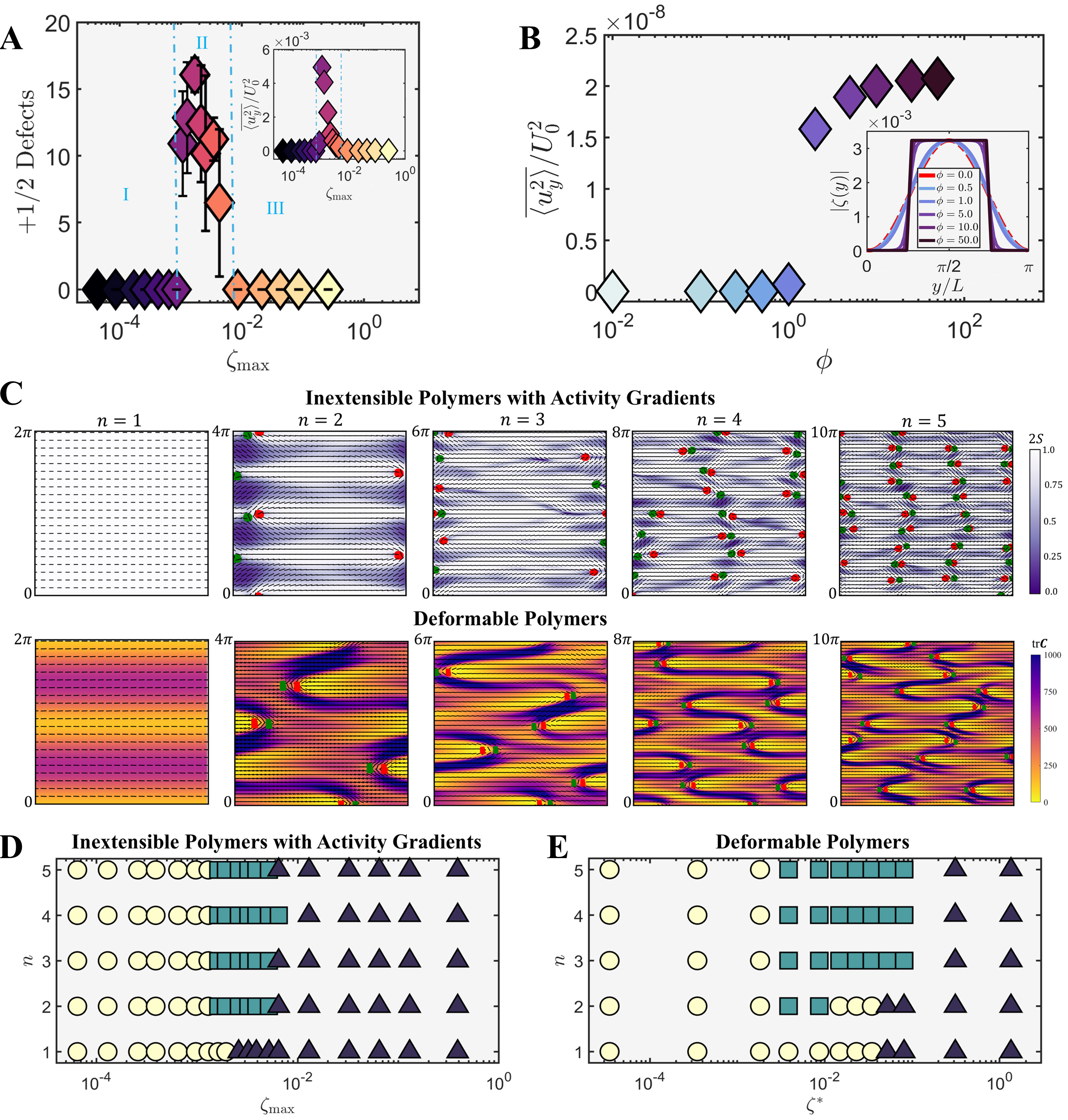}%width=0.95\textwidth
	\caption{\textbf{Comparing inextensible polymers (active nematics) with activity gradients to polymers that can stretch.}
    (\textbf{A}) Number of $+1/2$ defects for different $\zeta_{\max}$ in the statistical steady-state. The inset shows the corresponding temporally averaged mean secondary-flow strength, $\langle\overline{u_y^2}\rangle/U^2$. 
    (\textbf{B}) To demonstrate the role of $\nabla\zeta$, we conduct additional simulations at $\zeta_{\max}=0.006$ and control the steepness $\phi$ (Materials and Methods) of the imposed activity profile. The transition into secondary flows is enhanced as the sinusoidal activity profile transitions into a sharper ``step-like" response.  The inset shows the activity profile $|\zeta(y)|$ across half a forcing wavelength for different $\phi$ values. 
%     \ved{Avg. quanitity def's should now be consistent \& correct}.
   %  Notably, the transverse instability is associated with the formation of topological defects. 
    % (\textbf{D}) To demonstrate the role of $\nabla\zeta$, we conduct additional simulations at $\zeta_{\max}=0.006$ and control the steepness $\phi$ (Materials and Methods) of the imposed activity profile. The transition into secondary flows is enhanced as the sinusoidal activity profile transitions into a sharper "step-like" response.  The inset shows the activity profile $|\zeta(y)|$ across half a forcing wavelength for different $\phi$ values. %\ju{better as a linear scale following Sumesh's comment - the small $\phi$ values may be spurious velocities}
     (\textbf{C}) Contour plots for (top row) the degree of alignment measured by $S$ for active nematics with spatially varying activity and (bottom row) $\operatorname{tr(\mathsfbi{C})}$ for deformable polymers where the activity gradient is space and time dependent. In both cases, the director field $\bm{n}$ is superimposed with $+1/2$ (red) and $-1/2$ (green) topological defects. Snapshots are provided for different square domain sizes $n$. (\textbf{D}) and (\textbf{E}) Diagrams showing the final states identified across different domain sizes and activity levels for the active nematics and polymeric fluids, respectively. In both plots, three distinct dynamical regimes emerge. At low activity and small domain size, the system exhibits passive Kolmogorov flow with no defects (yellow circles). For moderate activity and sufficiently large domains, transverse flows develop, and defect pairs appear (green squares). At high activity, the system transitions into a flow-arrested state characterized by the absence of defects (purple triangles). }
     %\su{To go well with the section title E I suggest lets use 'Inextensible polymers with activity gradients' and 'Deformable polymers' as the figure titles. I would also get rid of the phrase 'state diagram' from B and C}} 
	\label{activitygradient}
\end{figure*}
 % , measures the degree of nematic order \ju{in the active nematic limit} \su{but otherwise contains information about how extended the polymers are and how aligned they are}).
 In each case in Fig.~\ref{fig2}A-C, there is a 
 %\cyanstrike{rapid} 
 sharp crossover between the force dominated ($\zeta^* < \mathcal{O}(1)$) and activity dominated ($\zeta^* > \mathcal{O}(1)$) regimes. Further, in each case there is a data collapse for varying  $Re$,  showing that the physics is controlled by a balance between the Kolmogorov forcing and the activity.

Figs.~\ref{fig2}D and E show the variation of the director field across the channel, and the corresponding forces which result in this behavior, in the flow dominated (left panels) and activity dominated (right panels) regimes. 
At low activities, the Kolmogorov forcing (red arrows in Figs.~\ref{fig2}E) dominates, and the scaled velocity magnitude is equal to $2/\pi$, as expected for a sinusoidal flow profile. The polymer configuration is independent of the activity, emphasizing the dominance of the imposed flow. For the choice of flow aligning parameter $\lambda=1$, polymers lie along the $x$-axis (Fig.~\ref{fig2}B and D) and the nematic order parameter averaged across the channel takes the constant value $\langle S \rangle =0.5$, corresponding to strong nematic ordering of stretched polymers (Fig.~\ref{fig2}C). 
%$\langle S \rangle =0.5$, corresponding to strong \bluestrike{polymer alignment}\su{nematic ordering of stretched polymers} (Fig.~\ref{fig2}C). 

%\cyanstrike{Once the activity is sufficiently large, the aligned polymers undergo the well-known active contractile instability \cite{VoituriezEPL,Voituriez,LSA_Ramaswamy,Edwards_2009} to give a splayed director configuration (Fig.~\ref{fig2}D), with the splay increasing with increasing activity (Fig.~\ref{fig2}A).}

With increasing activity, the aligned polymers exhibit a sharp cross-over to a splayed director configuration characteristic of the well-known contractile active instability \cite{VoituriezEPL,Voituriez,LSA_Ramaswamy,Edwards_2009} (Fig.~\ref{fig2}D), with the degree of splay increasing with  activity (Fig.~\ref{fig2}B). The resulting contractile stresses (blue arrows in Figs.~\ref{fig2}E) oppose the Kolmogorov forcing and the net velocity in the channel starts to decrease (Fig.~\ref{fig2}A). As the flow decreases, 
%it is less able to \su{stretch and} align the polymers so 
it becomes less effective at stretching and aligning the polymers, so the nematic order parameter, $\langle S \rangle$, also decreases (Fig.~\ref{fig2}C), approximately %\cyanstrike{linearly with} 
inversely proportional to $\zeta^*$. 
%Thus, in the limit that activity dominates the Kolmogorov flow, there is cancellation between the active and Kolmogorov forces leading to reduced flow along $x$ and a jamming state emerges.
Thus, in the limit where activity dominates the Kolmogorov flow, cancellation between the active and Kolmogorov forces leads to strong suppression of the flow along $x$. %\vedcomment{This suppression becomes more pronounced with increasing $Pe$, as seen from the decreasing streamwise velocity magnitude in Fig.~\ref{phase_diagram}B.}

\subsection{\textbf{Increasing the activity at high Pe}}

We next consider the effect of increasing the activity at higher P\'eclet  numbers where the system becomes unstable to flows in the $y$-direction over an intermediate range of activity. This 2D instability requires a sufficiently large active driving force $\bnabla \bcdot \left[\zeta^\ast \left(\mathsfbi{C}-\mathsfbi{I}\right)\right]$ in the equations of motion~(\ref{nondimu}), and hence, is favored by both increasing activity and increasing gradients in the polymer conformation tensor. The latter arise primarily from polymer extension induced by flow gradients and are therefore more readily generated at large $Pe$. In the flow-suppressed regime, however, the 1D instability opposes the Kolmogorov flow, reducing flow gradients and, hence, gradients in the polymer conformation tensor. Consequently, the 2D instability is suppressed throughout this regime for all $Pe$, despite the large activity.

When the system is unstable to 2D flows, a  lattice of velocity rolls is formed (Fig.~\ref{highPe}A). The mechanism behind this is illustrated schematically in  Fig.~\ref{highPe}B. Forces from the stress gradients produce transverse flows which, together with the constraint of incompressibility, create velocity vortices. In Fig.~\ref{highPe}C, we show that the vortices are associated with the appearance of the long-lived arrowhead structures in $\operatorname{tr(\mathsfbi{C})}$, which are well-documented features of elastic turbulence \cite{Morozov_Coherent,Lewy_Kerswell_2025,Page_PRL,KF_Waves}, as shown in 
Fig.~\ref{highPe}A.

 Given the connection between the descriptions of active and elastic systems, a natural question is whether the director field and topological defects, which are important concepts in active nematic physics, might be used to help characterize polymeric flows. Therefore, in Fig.~\ref{highPe}C, we plot the additional information contained in $\mathsfbi{C}$. This includes the associated director field, defined by the principal eigenvector of $\mathsfbi{C^*}$, as well as the locations of $\pm 1/2$ topological defects identified in this field.  Note that $\mathsfbi{C}$ and $\mathsfbi{C^*}$ share the same eigenvectors, and the director field therefore indicates the direction of elongated polymers.
%MAYBE ADD travelling waves propagating in opposing directions. They are quite well structured and originate from the well-known center-mode instability}. What is new here, is that arrowheads are marked by a pair of ±1/2 defects, which appear to occur in critical regions of the flow.
Strikingly, each arrowhead is associated with a pair of topological defects in the director field (Fig.~\ref{highPe}C). Movies~S2 and S4 show that this association persists dynamically for both single- and double-arrowhead states, demonstrating that it is a robust feature of the coherent structures.
% \ju{need to explain movies more}
The defect pairs are located in the vicinity of, but do not coincide with, the flow stagnation points in the co-moving frame identified in \cite{MOROZOV_ARXIV}.

%\ju{this paragraph may be too repetitive} The three different flow behaviours are also seen in Fig.~\ref{fig1}D which shows the time series of the dimensionless spatially averaged kinetic energy $E(t)/E_0=\langle \mathbf{u}^2\rangle/U^2$ for different values of $\zeta^*$. For small values of activity the system converges to steady laminar Kolmogorov flow, whereas intermediate values  generate strong time-dependent oscillations in the kinetic energy, characteristic of elastic turbulence. Larger values of the activity lead to a jammed response where the flow field in the channel is close to zero. 

Further increasing the P\'eclet  number to $\mathcal{O}(10^3)$
% at $Wi \sim \mathcal{O}(10)$OK?} 
tends to destabilize the pattern of arrowheads, creating secondary filament structures \cite{KF_Waves} (Fig.~\ref{highPe}A and Movie S5).  However, long-lived arrowhead structures persist, albeit in a less ordered arrangement. The arrowheads 
continue to correspond to bound defects \cite{MOROZOV_PNAS}, and the disappearance and reappearance of arrowheads correspond to the annihilation and birth of defect pairs.

Fig.~\ref{highPe}F shows the number of $+1/2$ defects as a function of activity at $Pe = 250$. Defects form at intermediate activities, where they are associated with arrowheads and 2D  flows, while no defects are observed at low or high activities. These three activity regimes correspond to the three regimes identified in the phase diagram in Fig.~\ref{phase_diagram} and coincide with regimes of minimal and maximal flow resistance (see inset).

%but there are no defects at low or high activities.

%To investigate this non-monotonic dependence on activity (and thus on the $Wi$) in more detail, we first solve the polymer equations of motion~[Eq.~\eqref{nondimu} and Eq.~\eqref{nondimC}] in the active nematic limit by keeping $\operatorname{tr(\mathsfbi{C})}$ constant. We then explicitly impose spatial gradients in $\operatorname{tr(\mathsfbi{C})}$ that are fixed in time, representing spatially varying, time-independent activity, and finally compare these results to the solutions of the full polymer equations, where the activity evolves in both space and time.
%, then add spatial gradients in $\operatorname{tr(\mathsfbi{C})}$ that do not vary in time, corresponding to spatially varying, time-independent activity,  and then compare to the solutions of the full polymer equations.

%This is reminiscent of channel flows in an active nematic where there is an instability driving a transition to a flowing state above a threshold activity for extensile systems or contractile systems which are nematic in the passive state. We are, however, unaware of any work on  contractile systems with no passive nematic ordering where there is no dynamics without an imposed flow. 

\section{\textbf{Spatial activity gradients for inextensible polymers}}
\label{Sec:gradient}

Having established the mapping between polymeric and active-nematic hydrodynamics, we next ask whether the transverse instability observed in polymers can be induced in an active-nematic system. Equation~\eqref{stressmap} relates the activity coefficient directly to $\operatorname{tr}(\mathsfbi{C})$, while our results above show that spatial variations in $\operatorname{tr}(\mathsfbi{C})$ play a key role in the transverse instability. We therefore hypothesize that spatially varying activity can induce a similar instability behavior in inextensible polymers, leading to defect pairs and associated velocity vortices.

 % in place of the arrowhead structures observed in the deformable polymer system
%\subsection{Constant activity gradients}%Indeed we found no evidence of a  2D instability in simulations in where the activity was held constant.

%Previous numerical work \cite{Morozov_Coherent,MOROZOV_ARXIV,King_Poole_Fonte_Lind_2025} has demonstrated that elastic turbulence is marked by thin regions of highly deformed polymers, reflected in localized variations of  $\operatorname{tr(\mathsfbi{C})}$, which in the active nematic limit, corresponds to changes in activity.
%Therefore, to investigate a possible role of activity gradients in driving the transverse instability, 
To test this, we impose a nematic activity gradient along $y$ according to a modified hyperbolic tangent profile for the activity coefficient,
\begin{equation}
\zeta(y) = \frac{\zeta_{\max}+\zeta_{\min}}{2} 
- \frac{\zeta_{\max}-\zeta_{\min}}{2} 
\frac{\tanh\!\big(\phi \cos(4 \pi y)\big)}{\max\big|\tanh\!\big(\phi \cos(4 \pi y)\big)\big|},
\end{equation}
%\ju{check $\zeta$ or $\zeta^*$}\sucomment{We can safely replace with $\zeta^*$ as its linear in $\zeta$, but Vedad to ensure consistency with Fig.3D} \ved{note: you can write with $\zeta^*$, I refrained from doing this, as the dimensionless values are 2-3 order of mag. larger than corresponding $\zeta^*$ value for polymers...}
where $\phi \to 0$ recovers a pure sinusoidal profile, while finite values of $\phi$ produce progressively steeper variations in $\zeta(y)$ (inset of Fig.~\ref{activitygradient}B). 
%\bluestrike{This situation is simpler than the full polymer simulations because, although the activity or equivalently the polymer extension, $\operatorname{tr(\mathsfbi{C})}$, varies in space, it is constant in time.} 
Note that the activity has half the wavelength of the Kolmogorov flow profile so that there is a maximum in the activity  to coincide with positions of maximum strain rate in the flow. Such activity variation is natural in polymeric fluids when subjected to the Kolmogorov flow, as $\operatorname{tr(\mathsfbi{C})}$ is maximal (minimal) where the velocity gradient is maximal (minimal).

In Fig.~\ref{activitygradient}A, we plot the number of defects as a function of $\zeta_{max}$. The inset of the figure shows the corresponding normalized, time-averaged, mean-square secondary flow for fixed $\phi=0$. We can identify the same three regions as  for the polymeric systems.
%\bluestrike{for the full polymer equations (compare Fig~\ref{highPe}F)}}. 
Region I corresponds to streamwise Kolmogorov flows, with no topological defects; Region II corresponds to the development of vortices and topological defects, and Region III is a flow-arrested state with no topological defects. 
In Fig.~\ref{activitygradient}B we vary the steepness of the activity profile for fixed $\zeta_{min}, \zeta_{max}$, demonstrating that the activity gradient is an important control parameter for the transverse instability -- the steeper the profile, the stronger the resulting flows along $y$.

 %\subsection{\textbf{Comparing imposed activity gradients t with activity gradients to deformable polymers}} 
% \ju{Rich paper} has shown 

Recent simulations of elastic turbulence under Kolmogorov forcing have shown that the transverse instability and resulting flow structures depend on the size of the simulation domain \cite{Lewy_Kerswell_2025}. We use this observation as a basis for a further comparison. Snapshots comparing the two cases in square simulation boxes, for different values of $n$, the number of Kolmogorov wavelengths across the box, are shown in Fig.~\ref{activitygradient}C  (see Fig.~S1 in the SI for the corresponding flow fields). In both systems, topological defects and associated secondary flows emerge only in sufficiently large domains, i.e., $n\geq2$. The pattern of topological defects is qualitatively similar in the two cases (see Movies S2 and S6), suggesting that the same instability is responsible for the transverse flows. However, as expected, the details of the pattern varies as $\operatorname{tr(\mathsfbi{C})}$ is allowed to vary with time as shown in the snapshots in the lower panel, which correspond to the polymeric system. This leads to the formation of the coherent structures which comprise arrowheads, distinctive regions of large $\operatorname{tr(\mathsfbi{C})}$ associated with a bound defect pair.

As a second way to compare the two systems, we present phase plots showing the long-time dynamical states as $n$ is varied. Fig.~\ref{activitygradient}D shows results for the active nematic as a function of $\zeta_{max}$ for the purely sinusoidal profile (i.e., $\phi\rightarrow0$) with fixed $\zeta_{min}=-1.3\times10^{-5}$. Fig.~\ref{activitygradient}E shows a similar plot for polymers, where now the control parameter is the effective activity $\zeta^*$. There is a striking similarity between the two cases: the transverse instability leads to a vortex lattice and topological defects at intermediate activities, sandwiched between the Kolmogorov flow state and the flow-arrested state, as $n$ is increased.

%In Fig.~\ref{fig3}D,  we plot the normalised, averaged mean square velocity across the channel as a function of $\phi$, which controls the magnitude of the activity gradients, at a time ??. This clearly shows that the transverse flows are favoured by steeper gradients in activity giving further evidence that variations in activity needed for the transverse instability to develop.
%We performed simulations across $\phi \in [0.01, 0.1, 0.25, 0.5, 1, 2, 5, 10, 25, 50]$ to systematically investigate the effect of gradient steepness on the transverse instability.
%Fig.~\ref{fig3} shows that once a threshold value in the amplitude of the imposed activity variation is reached, a transverse instability drives the system to a vortex state? Fig.~\ref{fig3} presents a series of snapshots showing the development of the instability.\\

% \subsection*{Solving the complete polymer equations}

\section{Discussion}
\label{discussion}
The central outcome of this work is that low–Reynolds-number flows in driven polymeric fluids and contractile active nematics can be investigated within a unified continuum framework.
% \vedstrike{The primary aim of this paper has been to describe the connection between low Reynolds number turbulence in driven dilute polymer suspensions and contractile active nematics.} 
We show that both systems correspond to the same equations of motion if the active nematic is isotropic in the absence of activity and is deformable. More broadly, this framework places active nematics, deformable active nematics, and polymeric fluids within a common description, allowing their dynamical instabilities and coherent structures to be understood in relation to one another. We present a unified phase diagram encompassing both flow suppression and the transverse instability leading to secondary flows. Crucially, the latter emerges from the full polymer dynamics, in which spatial variations of polymer extension play a role analogous to that of spatially varying activity in active nematics.

This correspondence suggests that topological defects and the nematic director field, which are widely used to characterize active flows, may provide new insights into polymeric fluid flows and elastic turbulence. In particular, we show that the arrowheads that characterize turbulence in polymeric fluids can be interpreted as long-lived coherent states formed around topological defect pairs, which appear at positions that correspond to large splay deformations in the director field. 

Conversely, the polymer literature will provide promising yet unexplored avenues for studying active nematics. 
Because contractile active nematics, which are isotropic in the passive limit, have no active dynamics in the absence of an imposed flow, they have received little attention so far. Motivated by our mapping, we studied the behavior of these systems in the Kolmogorov flow field. We found that when imposed flow dominates activity, the nematogens are aligned by the flow. However, at sufficiently large activities, they undergo the well-known active instability to a splay configuration, which creates stresses that oppose the Kolmogorov forcing. As a result, the velocity, and hence the nematic order, decreases, leading to a flow-suppressed state. 

A similar instability, from no flow to spontaneous laminar flow, driven by the usual active nematic bend or splay instability, is well known in channel-confined extensile active nematics \cite{VoituriezEPL,Voituriez,LSA_Ramaswamy,Edwards_2009}, and in contractile systems with thermodynamically-imposed nematic ordering. At higher activity levels in these models, the flow transitions to a line of flow vortices aligned along the channel \cite{Shendruk17}. In the parameter regime explored here for the active paranematic under Kolmogorov flow, we found no evidence of such a transition with increasing, constant, activity. We explain this by noting that activity gradients must be sufficiently strong to drive a transition to the vortex state, but before this happens, the activity becomes large enough to initiate a 1D instability which leads to flow-suppression and reduces the ordering and flow-alignment of the nematic director field. %\rv{!!!!This might not be entirely correct based on what we know now from linear stability analysis ... but we can keep this here to avoid complications ... or be a bit more cautious e.g. by saying that "in the parameter space explore in this paper ..."!!!!!}

To drive a transverse instability and form secondary flow structures, we needed to either add activity gradients along the $y$-axis in the active nematic simulations or solve the full polymer equations. In both cases, this resulted in vortex formation and the production of topological defects for intermediate values of the activity.  Moreover, the results for the two models depended in the same distinctive way on the size of the simulated domain.
%simulation box. 
These similarities suggest that the center-mode instability {\cite{Lewy_Kerswell_2025}, implicated in initiating the transverse flow in polymers, may also be relevant to the transition from laminar to vortical flow in active nematics, an interesting direction for future research. By contrast, arrowheads formed only when the activity variation had the time dependence encoded in the dynamical equation for the polymer extension, $\operatorname{tr(\mathsfbi{C})}$.

In formulating the mapping, we ignored the elastic backflow stress term in the equations of active nematics which results from relaxation of the director field.
%that appears in the equations of active nematics \su{due to the following reason. The backflow stresses are due to the director field relaxing}. 
To leading order, this is given by $\lambda A \mathsfbi{Q}$ which is negligible compared to the active stress
when $A$ is small or equivalently when $\operatorname{tr}(\mathsfbi{C})$ is large. The polymeric stress is also a backflow term, predominantly caused by stretched polymers relaxing their shape, and thus  $\propto \operatorname{tr}(\mathsfbi{C})$, which typically increases
with flow strength, leading to large polymeric stresses. The order-of-magnitude comparability between the polymeric stress and the dominating active stress term is the
fundamental reason behind Eq.~\eqref{stressmap}.

There are many directions in which to further investigate the polymer -- active nematic mapping. For example, confluent epithelial cell layers and tissues, which are deformable active nematics \cite{Hernandez_Marchetti}, can be modeled using the polymer equations of motion. Polymer dynamics in the absence of an imposed flow can be initiated by using an effective extensile activity
or a free energy term that forces the polymers to remain in a state with $\mathsfbi{C}\neq\mathsfbi{I}$. This provides a continuum description of active polymers which have recently received considerable attention in the literature \cite{Zhang25, levanon2025active}.%\su{citation added} 

While we focus on Kolmogorov flow, we expect similar behavior in flows, such as Poiseuille flow, where there are shear gradients. Moreover, here we considered the contractile, high-temperature limit where there is no active turbulence in the absence of flow and the flow-suppression transition pre-empts turbulence for driven active nematics. Therefore an interesting alternative would be to compare extensile, high-temperature active nematics, which do show active turbulence, to active polymers. It would also be interesting to move beyond low Reynolds number and consider deformable active nematics in the elasto-inertial regime, to consider polymeric turbulence in 3D in the context of nematic disclination lines, and to study whether more complex representations of polymer rheology can be cast into an active nematic framework. 

Lastly, in contrast to polymer molecules, many experimentally studied active systems are composed of colloidal-scale constituents, which allow direct visualization of the evolving microstructure. Our mapping between polymeric and active nematic systems therefore offers a potential route to testing theoretical and computational findings—particularly those concerning microstructural evolution—that are difficult to access experimentally in dilute polymer solutions.

\section{Materials and Methods}
\subsection{Numerical Method}
Equations~(\ref{nondimu}) and (\ref{nondimC}) are solved using a numerical solver developed in-house, comprising of the lattice Boltzmann (LB) method coupled with a high-order finite-difference scheme (see \cite{DZANIC2022105280}), which was applied in our previous investigations of elastic turbulence flow problems \cite{vedad_jfm,Dzanic_PRE, Dzanic_PHYS_FLUID, DzanicFrom2024,from2025, Vedad_PNAS_Nexus}. The hydrodynamic equations are solved using a single-relaxation-time (SRT) LB method on a standard D2Q9 lattice. The polymer field is resolved using a high-resolution finite-difference scheme, where spatial gradients are solved using a second-order central difference scheme, while a fourth-order Runge-Kutta scheme is applied for the temporal evolution. Careful attention is given to the advection term in Eq.~\eqref{nondimC}, which is discretized using the high-resolution Kurganov-Tadmor scheme \cite{KT_SCHEME}. To prevent the rapid growth of Hadamard instabilities \cite{SURESHKUMAR199553}, we further conserve the symmetric positive-definite (SPD) properties of the conformation tensor $\mathsfbi{C} \gneq 0$ by applying the Cholesky-log decomposition \cite{VAITHIANATHAN20031}. Full numerical details of the solver can be found in Dzanic \textit{et al.} \cite{DZANIC2022105280}. 

In terms of simulation parameters, we use LB units, where the discrete space and time steps are chosen as unity, and all quantities can be expressed in dimensionless form. Both rigid and extensible polymers are assumed to be flow-aligning particles with $\lambda=1$, which produce contractile force dipoles, hence $\zeta<0$. Following previous work \cite{Vedad_PNAS_Nexus,from2025}, we discretize the domain using $L=128/2\pi$, with further refinement to $L=256/2\pi$ yielding indistinguishable results and confirming numerical convergence. Unless specified otherwise, all simulations are performed with the parameters, $n=4$, $Re=1$, $Re_F=1$, $Wi\approx20$, and $Pe\approx250$. %\redstrike{Alternative parameter choices are provided in the online supplementary materials. }

Simulations are commenced under an initially isotropic, quiescent flow state where $\mathsfbi{C}=\mathsfbi{I}$ and $\mathbf{u}=0$. To simulate elastic turbulence, a small perturbation $\boldsymbol{\delta}$, a random Gaussian (normally distributed) noise in the order of $\mathcal{O}(10^{-6})$, is introduced to the initial velocity field $\mathbf{u}(t=0,\boldsymbol{x})+\boldsymbol{\delta}$, thereby seeding the instability. Fully periodic boundary conditions are employed in both spatial directions to model the Kolmogorov flow configuration. 

\subsection{Controlling Activity Gradients}
To investigate the role of spatial activity gradients $\nabla\zeta$ on the instability mechanism, we define a sinusoidal activity, which is constant in time,
\begin{equation}
\zeta(y)=\frac{\left(\zeta_{\max}+\zeta_{\min}\right)}{2} - \frac{\left(\zeta_{\max}-\zeta_{\min}\right)}{2}\cos\left(4\pi y\right),
\end{equation}
which periodically varies between $\zeta_{\min}$ and $\zeta_{\max}$, with a wavelength half that of the Kolmogorov forcing. Throughout, we fix the minimum activity level at $\zeta_{\min} = -1.3\times 10^{-5}$. In doing so, the contractile stress contribution from rigid polymers (i.e., $\operatorname{tr}(\mathsfbi{C})$ held constant) becomes $\sigma_p\approx\zeta(y)\mathsfbi{C}^*$.

To elucidate the role of activity gradients in driving the transverse instability reported in the results, we systematically control the steepness of $\nabla\zeta$ using the modified hyperbolic tangent profile,
\begin{equation}
\zeta(y) = \frac{\zeta_{\max}+\zeta_{\min}}{2} 
- \frac{\zeta_{\max}-\zeta_{\min}}{2} 
\frac{\tanh\!\big(\phi \cos(4\pi y)\big)}{\max\big|\tanh\!\big(\phi \cos(4\pi y\big)\big|},
\end{equation}
where $\phi \to 0$ recovers the pure sinusoidal profile, while finite values of $\phi$ produce a progressively step-like variation in $\zeta(y)$. We perform simulations across $\phi \in [0.01, 0.1, 0.25, 0.5, 1, 2, 5, 10, 25, 50]$ to systematically investigate the effect of gradient steepness on the transverse instability.

%\sucomment{To cleanup}5. Should $\zeta$ be $\zeta^*$ in this figure.(3)
%Probably. Currently $\zeta_max$ is in dimensional units. Making it dimensionless (i.e., $\zeta^*$) would give something like $\zeta_max = 0.005 ---> \zeta^* =  13.5$, as an example. This is quite high when compared to the values for $\zeta^*$ in Fig 1 for extensible polymers. Perhaps discrepancies arise due to trC. For instance trC_avg ~ 1000, hence $\zeta^*/trC_avg = 13.5/1000 = O(10^-2)$, something more reasonable I suppose.

\section*{Acknowledgments}
We thank Rich Kerswell for fruitful discussions. V.~Dzanic acknowledges support from the Australian Research Council Discovery Project (Grant No.~DP230102229) and the high-performance computing facilities at QUT. SPT acknowledges funding from the Royal Society and the Wolfson Foundation through a Royal Society Wolfson Fellowship and support from the Department of Science and Technology, India, under research grant CRG/2023/000169. JMY and RV acknowledge support from the ERC Advanced Grant ActBio (funded as UKRI Frontier Research Grant EP/Y033981/1). 

% \showacknow{} % Display the acknowledgments section

% \bibsplit[2]
%Use \bibsplit to split the references from the body of the text. Value "[2]" represents the number of reference in the left column (Note: Please avoid single column figures & tables on this page.)

% Bibliography

\bibliography{PNAS_BIB}

\end{document}